\documentstyle[aps]{revtex}

\def\vecF{\mbox{\boldmath$F$}}
\def\vecJ{\mbox{\boldmath$J$}}
\def\vX{\mbox{\boldmath$X$}}
\def\vB{\mbox{\boldmath$B$}}

\def\vecPi{\mbox{\boldmath$\Pi$}}
\def\vx{\mbox{\boldmath$x$}}
\def\vy{\mbox{\boldmath$y$}}

\def\vnabla{\mbox{\boldmath$\nabla$}}

\begin{document}

\title{Mesoscopic Nonequilibrium Thermodynamics of Single Macromolecules 
and Dynamic Entropy-Energy Compensation}

\author{Hong Qian}

\address{Department of Applied Mathematics\\
University of Washington, Seattle, WA 98195, U.S.A.\\
qian@amath.washington.edu} 

\date{\today}

\maketitle

\begin{abstract}
We introduce axiomatically a complete thermodynamic formalism 
for a single macromolecule, either with or without 
detailed balance, in an isothermal ambient fluid based on its 
stochastic dynamics. With detailed balance, the novel theory 
yields mesoscopic, nonequilibrium generalizations for entropy 
($\Upsilon_t$) and free energy ($\Psi_t$) of the macromolecule. 
$\Upsilon_t$ and $\Psi_t$ fluctuate. Expectation 
$(d/dt)E\left[\Psi_t\right]$ $\le 0$, ``='' holds if and 
only if the macromolecule is at thermal equilibrium, in which
we show $\Upsilon_t$ still fluctuates but $\Psi_t$ is a constant.  
The entropy fluctuation {\it a l\`{a}} L.D. Landau, 
$E\left[(\Delta\Upsilon_t)^2\right]$, precisely matches
the fluctuation in the internal energy, which 
in turn equals the fluctuation in heat dissipation.  As 
a generalization of Clausius' classic result, the dynamic 
fluctuations in the entropy and energy of the
macromolecule are exactly compensated at
thermal equilibrium.  For systems with detailed balance,
Helmholtz free energy is shown to 
be the potential of Onsager's thermodynamic force.
\end{abstract}

\vskip 0.3cm


\vskip 0.5in

	In recent years, the stochastic model for overdamped 
Newtonian Brownian dynamics in a force field,
\begin{equation}
       \Xi d\vX_t = \vecF(\vX_t)dt + \Gamma d\vB_t,
\label{NLSDE}
\end{equation}
has found an increasing number of applications to a host of 
macromolecular processes in equilibrium and more importantly
nonequilibrium steady-state (NESS) \cite{Q1,Q2}.  These
models are generalizations of the classical theory of 
polymer dynamics in which $\vecF$ usually is conservative 
\cite{DE}.  In addition to obtaining the stochastic dynamics 
from the model, however, it becomes clear 
to us that an axiomatic isothermal thermodynamic formalism can 
be developed based solely on the stochastic differential equation 
(\ref{NLSDE}) in which $\vX_t$ represents the coordinates of the 
``atoms'' in the macromolecule, $\Xi$ is a constant, positive 
definite hydrodynamic interaction matrix, $\vecF$ is the force
not necessarily conservative \cite{QPRL98}, and $\Gamma d\vB_t$ is a 
white noise representing the collisions between the 
macromolecule and the solvent: $\Gamma\Gamma^T$ = $2k_BT\Xi$ 
according to Einstein's relation and defines the temperature of
the isothermal system.  The solution to (\ref{NLSDE}), $\vX_t$, 
is stochastic, whose probability density function satisfies the 
Fokker-Planck equation
\begin{equation}
    \frac{\partial P}{\partial t} =
                \vnabla\cdot\left(\frac{1}{2}A\vnabla P
                                        - \Xi^{-1}\vecF(\vx)P\right),
  		\hspace{0.5cm}
                (A = \Xi^{-1}\Gamma\Gamma^T\Xi^{-T} = 2k_BT \Xi^{-1}),  
\label{NLFPE}
\end{equation}
where $P(\vx,t)$ is the probability of the macromolecule being in 
conformation $\vx$ at time $t$: $\vX_t =\vx$.  In this paper, we show 
how a complete, mesoscopic thermodynamic theory can be formulated
based on (\ref{NLSDE}), which we call {\it stochastic macromolecular 
mechanics} (SM$_3$) \cite{Q2}.  We then apply this novel formalism 
to further elucidate a classic observation in the equilibrium
thermodynamics of proteins: the dynamic origin of entropy-enthalpy 
compensation \cite{Q4}.  Naturally, the enthalpy under constant 
pressure is equivalent to the internal energy in our formalism.
The significance of the nonequilibrium 
steady-state obtained from this formalism \cite{Q3} and 
its applications to free energy transduction in biological 
macromolecules, e.g., a protein molecule converting chemical
potential into mechanical work, have been discussed elsewhere 
\cite{Q1,Q2,QPRL98}.  

	It is important to point out that even though there is already
a vast literature on stochastic models based on Fokker-Planck
equation (\ref{NLFPE}) \cite{DE,Hone}, it is not known that this 
approach rigorously encompasses a comprehensive statistical 
thermodynamics.  Furthermore, it is not known whether a
thermodynamics exists for individual macromolecules. 
In this paper, we demonstrate the logical relation between models 
based on equations (\ref{NLSDE}) and (\ref{NLFPE}) and the principles 
of equilibrium and nonequilibrium thermodynamics.  A complete 
statistical thermodynamics for isothermal systems in equilibrium, 
nonequilibrium transient and NESS, as an complementary and 
alternative to Boltzmann's approach, emerges. 

\vskip 0.5cm

{\bf Stochastic Macromolecular Mechanics.}
Following Lebowitz and Spohn \cite{LS,Q3}, we first introduce 
the instantaneous heat dissipation $W_t$:
\begin{equation}
        dW_t \equiv \vecF(\vX_t)\circ d\vX_t 
        = \vecF(\vX_t)\cdot d\vX_t 
		+ \frac{1}{2}d\vX_t \vnabla\vecF d\vX_t
\label{dW}
\end{equation}
where $\circ$ and $\cdot$ denote integrations in the Stratonovich 
and Ito sense, respectively \cite{Ok}.  According to this definition, 
the heat dissipation, $dW_t$, is equal to the work done by the system: 
the product of force $\vecF$ and displacement $d\vX$. This is the 
law of energy conservation.  Using the expression (\ref{NLSDE}) 
it is easy to show that the mean rate of the heat dissipation 
(HDR, $h_d$)
\cite{LS,Q3}
\begin{equation}
        h_d = \frac{d}{dt}E\left[W_t\right]
              = \int \vecF(\vx)\cdot \vecJ(\vx,t) d\vx
\end{equation}
where 
\begin{equation}
    \vecJ(\vx,t) = -\frac{1}{2}A\vnabla P(\vx,t)
		+\Xi^{-1}\vecF(\vx)P(\vx,t) 
\end{equation}
is the probability flux in (\ref{NLFPE}):
$\partial P(\vx,t)/\partial t$ = $-\vnabla\cdot\vecJ$.
In mechanical terms, the mean heat dissipation rate is the 
product of the molecular force and the flux.  

	The dynamic of $W_t$ is itself Brownian-motion like
with $h_d$ as its mean rate \cite{Q3}.  Its fluctuations can be
characterized by a ``heat diffusion coefficient'':
\begin{equation}
     E\left[\frac{(dW_t)^2}{dt}\right]
        = 2k_BT\int \vecF(\vx)\Xi^{-1}\vecF(\vx)P(\vx,t) d\vx,
\end{equation}
which has a dimension of [energy$^2$]/[time].  The 
statistical properties of $W_t$ have been extensively 
explored in connection to the Gallavotti-Cohen symmetry
and fluctuation theorem \cite{G,LS,Q3}.

        The Onsager's thermodynamic force is different from
the mechanical force $\vecF(\vx)$ \cite{On}.  In terms of
(\ref{NLFPE}), we introduce a second thermodynamic quantity,
$\vecPi(\vx,t)$, the thermodynamic force \cite{FN1}
\begin{equation}
     \vecPi(\vx,t)\equiv\vecF(\vx)
		-\frac{1}{2}\Xi A\vnabla\ln P(\vx,t),
\label{Oforce}
\end{equation}
and a third thermodynamic quantity: the entropy 
$S$ according to the well-known formula:
\begin{equation}
           S = -k_B\int P(\vx,t)\ln P(\vx,t) d\vx
\label{entropy}
\end{equation}
where $P(\vx,t)$ is the solution to (\ref{NLFPE}).  In terms of 
(\ref{Oforce}) and (\ref{entropy}) we have the increase of 
the entropy at constant temperature $T$
\begin{eqnarray}
  T\dot{S} &=& k_BT\int (\ln P + 1)\nabla\cdot\vecJ\
         dx    \nonumber \\
    &=& -\int (k_BT\nabla \ln P-\vecF)\cdot\vecJ\ dx
        -\int \vecF\cdot\vecJ\ dx
                \nonumber \\
    &=& \int \vecPi\cdot\vecJ dx
                        - h_d
                        \label{depr} \\
    &=& e_p - h_d.    \nonumber
\end{eqnarray}
in which we identify, following Onsager, $\int\vecPi\cdot\vecJ d\vx$ 
with the {\it entropy production rate} (EPR, $e_p$) \cite{Sch}.  
It can also be rewritten as 
\begin{equation}
    e_p = 
    \int \left(\frac{1}{2}A\vnabla\ln P(\vx,t)
		-\Xi^{-1}\vecF(\vx)\right)^T
   \Xi\left(\frac{1}{2}A\vnabla\ln P(\vx,t)
		-\Xi^{-1}\vecF(\vx)\right)P(\vx,t)d\vx.
\label{epr}
\end{equation}
which is always non-negative.  This is the second law of thermodynamics.  
(\ref{depr}) is valid for all isothermal Brownian dynamical systems, with 
or without detailed balance, in stationary state and in transit process.  
It encompasses both the first and second laws of thermodynamics. 
In fact, it makes the second law quantitative by providing a rate 
for entropy increase.   In a time independent stationary state, the 
$\dot{S} = 0$ in (\ref{depr}), and the entropy production $e_p$ is 
balanced by the heat dissipation $h_d$.  This is the general case 
for an isothermal NESS.  

	Eq. \ref{epr} also indicates that $e_p$ equals zero if and only 
if $\vecF$ = $(\Xi A/2)\vnabla\ln P$ = $\vnabla\ln P/k_BT$.  
That is the force field $\vecF$ has to be conservative with an
internal potential energy: $\vecF$ = $-\vnabla U$.  For system 
with the potential, also known as detailed balance \cite{Risken}, 
the stationary solution to (\ref{NLFPE}) is $P =Z^{-1}e^{-U/k_BT}$ 
where the normalization factor $Z$ is the partition function in 
Gibbsian equilibrium statistical mechanics (isothermal canonical
ensemble).  It can be mathematically shown that 
$e_p$ = 0 is a sufficient and necessary condition for 
the stationary stochastic process $\vX_t$ defined by 
(\ref{NLSDE}) being time reversible \cite{Q5}.  Therefore, 
time reversibility, detailed balance, and zero entropy production
are equivalent with an equilibrium.\cite{FN2}

	For systems satisfying the potential condition $\vecF$ 
= $-\vnabla U$, the thermodynamic force $\vecPi$ also has a 
potential: $\vecPi$ = $-\vnabla\Psi$ where $\Psi(x)$ = $U(\vx)+k_BT\ln P$. 
We note that the expectation of $\Psi_t$ $\equiv$ $\Psi(\vX_t)$:
\begin{equation}
       E\left[\Psi_t\right] =  E\left[U(\vx)+ k_BT\ln P(\vx)\right]
             = \int P(\vx,t)U(\vx)d\vx-TS
\label{fe}
\end{equation}
which in fact is the Helmholtz free energy \cite{FN3}!  The first 
term in (\ref{fe}) is the mean internal energy.  More importantly,
it is easy to show that 
\begin{equation}
        E\left[\dot{\Psi}_t\right] = h_d - T\dot{S}
			= -e_p \le 0.
\end{equation}
In an isothermal system, the Helmholtz free energy decreases and 
reaches its minimum at the equilibrium: $-k_BT\ln Z$.  This is 
precisely the statement of second law of thermodynamics for an 
isothermal system. 

	Finally, with the potential condition, we have
\begin{eqnarray}
       -dU(\vX_t) &=& -\vnabla U(\vX_t)\cdot d\vX_t 
		- \frac{1}{2}d\vX_t\cdot \vnabla\vnabla 
			U(\vX_t)\cdot d\vX_t 
\nonumber\\
                &=& \vecF(\vX_t)\cdot d\vX_t 
		 + \frac{1}{2}d\vX_t\cdot \vnabla 
				\vecF(\vX_t)\cdot d\vX_t.
\label{dU}
\end{eqnarray}
Comparing (\ref{dU}) with (\ref{dW}), we see that the heat 
dissipation $dW_t$ = $-dU(\vX_t)$, the internal energy fluctuation.  
Hence, $W_t$ = $-U(\vX_t)$ is stationary and its expectation and 
variance are the internal energy and heat capacity $(C_v)$ of a 
single macromolecule at thermal equilibrium.

\vskip 0.5cm

	{\bf The Matching Entropy and Energy Fluctuations.}  We now 
focus on systems with detailed balance.  The above thermodynamic formalism 
suggests the fluctuating $U_t$ $\equiv$ $U(\vX_t)$ as a mesoscopic, 
nonequilibrium generalization of internal energy of a macromolecule 
in an isothermal aqueous solution.  Then its expectation $E[U_t]$ = 
$\int U(\vx)P(\vx)d\vx$ which equals to the standard internal energy 
in thermodynamics.  Similarly in the same spirit, we can introduce 
fluctuating entropy $\Upsilon_t$ $\equiv$ $-k_B\ln P(\vX_t)$, which can be 
viewed as the mesoscopic, nonequilibrium generalization of entropy.  
It is important to point out that this definition is consistent 
with the Boltzmann's microscopic entropy based on the volume of
the phase space.  In our case $P(\vX_t)$ is the probabilistic measure 
of the phase space.  The expectation $E[\Upsilon_t]$ is the 
Gibbs entropy in Eq. \ref{entropy}.

	We now show that the mesoscopic generalizations immediately 
lead to an interesting thermodynamic result in equilibrium.  We note 
that while $U_t$ and $\Upsilon_t$ are fluctuating in equilibrium, their 
difference, 
\begin{equation}
        U_t - T\Upsilon_t = U(\vX_t) + k_BT\ln P(\vX_t) = -k_BT\ln Z    
\end{equation}
the equilibrium free energy, however, is not fluctuating.  We can 
further compute the fluctuations in the mesoscopic entropy and 
internal energy:
\begin{eqnarray}
     T^2E\left[(\Delta\Upsilon_t)^2\right] &=&
     E\left[(\Delta U_t)^2\right]
\nonumber\\
       &=& \int U^2(\vx)e^{-U(\vx)/k_BT}d\vx
	- \left(\int U(\vx)e^{-U(\vx)/k_BT}d\vx\right)^2
\nonumber
\\
       &=& k_BT^2 \frac{\partial}{\partial T}E[U_t] 
		= k_BT^2C_v.
\label{entflu}
\end{eqnarray}
Furthermore,
\begin{equation}
    TE\left[\Delta\Upsilon_t\Delta U_t\right] = 
	 E\left[(\Delta U_t)^2\right].
\end{equation}
Therefore, the fluctuations of the mesoscopic internal energy 
$U_t$ and entropy $\Upsilon_t$ are perfectly correlated. They 
compensate in the {\em dynamical fluctuationsi} of a macromolecule.

\vskip 0.5cm

	{\bf Entropy Fluctuation and Its Mesoscopic 
Interpretations.}
The relation between entropy fluctuation and heat capacity in
Eq. \ref{entflu} was known to L.D. Landau who also 
advocated the concept of entropy fluctuation \cite{LL}.  It 
has been a difficulty concept to many who consider
entropy to be a functional of the distribution \cite{Cooper}. 
Here we offer a more plausible interpretation of the 
concept based on our mesoscopic view from the previous
section. 

	Let's consider $N$ identical, independent macromolecules
in the aqueous solution, $(\vX_1,\vX_2,...,\vX_N)$, each with its own 
stochastic dynamic equation (\ref{NLSDE}).  The concentration
of the number of molecules in conformation $\vx$ can then be
defined as
\begin{equation}
      C_{x,t} = \sum_{j=1}^N \delta(\vx-\vX_j).
\label{concen}
\end{equation}
Note that since the $\vX$'s are stochastic, the concentration
$C_{x,t}$ fluctuates.  However, the classic statistical mechanics 
is only concerned with the {\it most probable $C_{x,t}$} since 
the relative fluctuation in $C_{x,t}$ is insignificant in the 
thermodynamic limit when $N$ $\rightarrow$ $\infty$.  
Nevertheless, the $C_{x,t}$ fluctuates.

	The macroscopic entropy is defined as a functional of 
the density function $C_{x,t}$.  Therefore, it fluctuates with 
the $C_{x,t}$.  We now show that this fluctuation is indeed the 
mesoscopic fluctuation introduced in the formalism for single 
macromolecules.  We consider a general thermodynamic quantity
\begin{equation}
       q =  \int Q(\vx)C_{x,t}d\vx.
\end{equation}
If one had neglected the fluctuation in $C_{x,t}$, a fluctuation in
$q$ would be inconceivable. 

	Noting Eq. \ref{concen}, the expectation of $q$ is readily
computed:
\begin{eqnarray*}
   E[q] &=& \int Q(\vx)d\vx\sum_{j=1}^N E\left[\delta(\vx-\vX_j)\right]
\\[5pt]
         &=& \int Q(\vx)d\vx \sum_{j=1} \int \delta(\vx-\vy_j)
			P(\vy_1,\vy_2,...,\vy_N)d\vy_1...d\vy_N 
\\[5pt]
         &=& N\int Q(\vx)P(\vx)d\vx = NE\left[Q(\vX_t)\right]
\end{eqnarray*}
where the joint probability for $\vX_1$, $\vX_2$, ...., $\vX_N$,
$P(\vx_1,\vx_2,...\vx_N)$ = $	P(\vx_1)P(\vx_2)...P(\vx_N)$ since the
macromolecules are assumed to be independent in the solution.
Similarly, the variance in $q$:
\begin{eqnarray*}
   VAR[q] &=& \sum_{j=1}^N VAR\left[\int Q(\vx)\delta(\vx-\vX_j)dx\right] 
\\[5pt]
         &=& \sum_{j=1}^N VAR\left[Q(\vX_j)\right] 
\\[5pt]
         &=& N VAR\left[Q(\vX_t)\right] = N VAR\left[Q_t\right] 
\end{eqnarray*}
in which $Q_t$ $\equiv$ $Q(\vX_t)$ \cite{FN5}.   Therefore, 
the fluctuations in $q$ due to fluctuating 
in the distribution function $C_{x,t}$,
is exactly the mesoscopic fluctuation we have introduced in
the previous section for single macromolecules.  The mesoscopic
view, however, clearly indicates a stochastic dynamic origin of 
these fluctuations.

\vskip 0.5cm
	
	{\bf Entropy and Heat.}  It is a classic result of 
Clausius that entropy change equals to heat dissipation in 
isothermal quasi-static processes.  Our result suggests that
both concepts can be generalized dynamically to isothermal
equilibrium, and their fluctuations are indeed equal, as 
we have shown.  In our formalism, the energy conservation is
instantaneous, hence $dW_t=-dU_t$.  The free energy, on the
other hand, has to be constant over the entire conformational 
space in an equilibrium, $\Psi(\vx)$ $\equiv$ $-k_BT\ln Z$,
while $\Upsilon(\vx)$ and $U(\vx)$ are not.  This is a
demonstration for the concept of {\it local equilibrium},
which is essential in the general theory on nonequilibrium 
thermodynamics \cite{GP}.   Therefore, $Td\Upsilon_t$ = 
$dU_t$ = $-dW_t$; and $dU_t/d\Upsilon_t$ = $T$, all 
{\it instantaneously}.

\vskip 0.5cm

	{\bf The Rate of $e_p$.} The $e_p$  defined 
in (\ref{epr}) is instantaneous and time-dependent in a 
nonequilibrium transient.  Thus, one can further compute its time 
derivative $(de_p/dt)$, the change in entropy production rate 
\cite{Hone}. With some algebra we have \cite{FN4}
\begin{equation}
   \frac{de_p}{dt} =
      -2k_BT\int \left(\vnabla\cdot\vecJ\right)^2 P^{-1}(\vx)d\vx
      - \int \left(\vecPi^T\Xi^{-1}\vecPi\right)
		\left(\nabla\cdot\vecJ\right) d\vx.
\label{deprdt}
\end{equation}
Near a NESS, $\vnabla\cdot\vecJ$ is small and the 
second term has the leading order.  However, with detailed balance
and near an equilibrium, $\vecPi$ is also small; hence the first 
term, which is negative, becomes the leading term.  Therefore, near 
an equilibrium, the $e_p$ {\it monotonically} approaches to zero.  
This is the Glansdorff-Prigogine's principle of minimal entropy 
production rate\cite{GP}.  Near a NESS, however, the second term 
dominates, and (\ref{deprdt}) is not necessarily negative.

\vskip 0.5cm

	{\bf Discussions and Summary.}  While our thermodynamic 
formalism is strictly for isothermal processes, it can be used to 
compare a system under different temperature.  In a thermal
equilibrium, as we have shown, the fluctuation in internal 
energy $U_t\equiv U(\vX_t)$ is directly related to the heat 
capacity: $C_v=dE[U_t]/dT$.  Such relations, if any, in
nonequilibrium steady-state have not been explored.
Furthermore, one can easily introduce an external force
into the Eq. \ref{NLSDE} to represent the pressure, and 
thus we expect a thermodynamic formalism for isobaric
systems can be developed in parallel.

	It is also noted that there is a difference 
between the mesoscopic energy $U_t\equiv U(\vX_t)$ and 
entropy $\Upsilon_t\equiv -k_B\ln P(\vX_t)$.  While the 
former can be computed along a stochastic trajectory, 
the latter can not until the probability distribution function
$P(\vx,t)$ is known.  This difference reflects the fundamental
difference between the two physical quantities, energy and
entropy.  The former is {\it local} while the latter is 
{\it non-local} due to circular balance:
one can not know the value of entropy of a state until
knowing how likely it occurrs in comparison with othe
states.  It is possible to formally express the 
$P(\vx,t)$ in temrs of a path integral \cite{Hak}.
Maes also suggested a space-time approach to the problem
\cite{Maes}. In a thermodynamic equilibrium, however, $P(\vx)$ 
can be determined locally up to a normalization factor due 
to detailed balance.  Hence, $d\Upsilon_t=TdU_t$.

	In summary,
\begin{eqnarray*}
      \Xi d\vX_t &=& \vecF(\vX_t)dt + \Gamma d\vB_t,
	\hspace{0.75in} (\textrm{conformational dynamics})
\\
      dW_t &=& \vecF(\vX_t)\circ d\vX_t,
	\hspace{1.in} (\textrm{heat dissipation})
\\
      \Upsilon_t &=& -k_B\ln P(\vX_t,t),
	\hspace{0.82in} (\textrm{entropy})
\\
      \vecPi_t &=&  \vecF(\vX_t) + T\nabla\Upsilon(\vX_t,t)
	\hspace{0.5in} (\textrm{thermodynamic force})
\end{eqnarray*}
is a complete set of equations which provides the stochastic 
dynamics of a macromolecule, its heat dissipation, its entropy 
(Boltzmann), and its thermodynamic driving force (Onsager).
With this set of equations, one can compute $h_d$ = $(d/dt)E[W_t]$
and $e_p$ ($\ge 0$) from entropy balance $T(d/dt)E[\Upsilon_t]$ = 
$e_p - h_d$. If a system is detail balanced, i.e., $\vecF$ = 
$-\vnabla U$. Then $W_t=U_t$ $\equiv$ $U(\vX_t)$, and $\vecPi$ = 
$-\vnabla\Psi(\vX_t)$ where $\Psi(\vX_t)$ $\equiv$ $\Psi_t$  
= $U_t-T\Upsilon_t$ is free energy:
$(d/dt)E[\Psi_t]$ =-$e_p$ $\le 0$ and reaches its minimum 
$-k_BT\ln Z$ at equilibrium.  In the equilibrium, the probability
distribution for $\vX_t$ is $Z^{-1}e^{-U(\vx)/k_BT}$, 
$U_t-T\Upsilon_t$ = $-k_BT\ln Z$, and $E[(\Delta\Upsilon_t)^2]$
= $(1/T^2)E[(\Delta U_t)^2]$ = $k_BC_v$, the heat capacity.

\acknowledgments

I thank Marty Cooksey, Bernard Deconinck, Shoudan Liang, Christian
Maes, and Kyung Kim for many helpful discussions.


\begin{references}
\bibitem{FN1}
The $\vecPi$ corresponds to the affinity in the master
equation formalism $A_{ij}=k_BT\ln(p_jq_{ji}/p_iq_{ij})$
where $\{q_{ij}\}$ is the transition rate and $p_i$ the
state $i$ probability. Similarly, the ``mechanical work''
$\int\vecF(X_t)\circ dX_t$ corresponds to 
$k_BT\ln(q_{ji}/q_{ij})$.  See \cite{Sch,Q3} for more details.
\bibitem{FN2}
In the literature, the notion of detailed balance has two 
closely related, but different meanings.  For a stationary
Markov process, it means the transition flux from $i$ to $j$
equals exactly the transition flux from $j$ to $i$; hence the
net flux is zero.  This notion can be mathematically shown to 
be equivalent to time-reversibility.  Stationarity does not
imply time-reversibility.   Detailed balance is also 
used as a property of a Markovian system which, when reaching
its stationary state, is time reversible with zero flux. 
The second notion can be mathematically shown to be equivalent 
to the potential condition.  By the laws of thermodynamics, 
an equilibrium has to have detailed balance; and a system 
without energy input has to be detail balanced.
\bibitem{FN5}
In terms of the probability theory, $\vX_k$ are independent, 
identially distributed random variables.  Hence $Q(\vX_k)$ 
are independent, identically distributed.  Therefore, the 
expectation and variance of the sum of $Q(\vX_k)$ is the sum of 
the expectation $E[Q(\vX_k)]$ and variance $VAR[Q(\vX_k)]$.
\end{references}
\end{document}